# SELECTIVE PRIVACY IN IOT SMART-FARMS FOR BATTERY-POWERED DEVICE LONGEVITY

**ABSTRACT**

This paper presents a payload security model to maintain the standards of TLS whilst removing obstacles associated with constrained devices and IoT network protocols. The domain of aquaponics as a smart-farming environment was used to test the novelty with ESP32 edge devices. The impact of privacy on message content illustrated the most appropriate security attribute configurations, followed by time and power analysis to calculate energy consumptions of each scenario. It was concluded that using this tailored payload security model rather than TLS was capable of extending the battery life of constrained devices by up to 81%, whilst maintaining TLS security standards, and applicable to various protocols.

## 1. INTRODUCTION

With multiple threats to food systems including climate change, population increase, increase of transport costs and so forth, the demand for food production will increase by 70% by the year 2050 (Demestichas, Peppes and Alexakis, 2020). In response to this emerging threat, agriculture is transforming; IoT-powered smart-farms enable autonomous and self-sufficient food production systems to operate remotely (Vermani, 2019). Amongst the multitude of smart-farm solutions available (Navarro, Costa and Pereira, 2020), aquaponics represents a symbiotic relationship between plants and fish - fish produce ammonia from which bacteria then creates nitrates. These nitrates are then pumped around the planted beds as a natural fertiliser, several times an hour, with monitored temperature, PH level, and other variables necessary for plant-specific farming. Such an environment exemplifies a Wireless Sensor Network (WSN), or a series of spatially dispersed and dedicated IoT devices (Czelusniak *et al.*, 2019). This paper presents mechanisms for reducing energy usage for securing IoT control and monitoring devices in an aquaponics-based smart farming application.

Food production is now a critical issue, and attacks towards smart-farms add to the myriad of existing IoT-oriented information attacks such as CCTV (Doshi, Apthorpe and Feamster, 2018), or even children's toys (Keymolen and Van der Hof, 2019). Specifically regarding food, smart city attacks have witnessed sabotage of restaurant freezers and food poisoning (Rashid *et al.*, 2020). With a general indication of attacker interest in IoT, and a future where the population is high and food supply low, threats towards food supplies may be imminent. With such traction in both information attack vectors and the rise of IoT-powered sustainability, security is a basic requirement.

The issues pertaining to security are that of a 'balancing act' (Kane *et al.*, 2020), where integration is unattractive for two main reasons. Firstly, the value of security is not perceived until loss (Gan and Heartfield, 2016), and so if the domain is considered to be of little value to attackers in general, or beyond the chances of being targeted because they are considered 'low impact', then security can be seen as adding costs without benefits. Secondly, the predefined ciphersuites available to replicate the de facto security standard Transport Layer Security (TLS), are tremendously draining on constrained devices. For this reason security is considered a challenge (Del-Valle-Soto *et al.*, 2019), as TLS needs infrastructure, energy and processing - three things that contradict the nature of constrained IoT devices and networks.

In addition to the tension between TLS and IoT, the HTTP protocol is not ideal for IoT applications. Lighter protocols such as MQTT, LoRaWAN and BLE have their own security designs - usually based on TLS, and accompanied by the same problems. Perhaps if a decentralised TLS security model existed to secure payloads rather than protocols with considerably less burden on battery life, then the notion of including security as a fundamental requirement would not be so intimidating.

This paper presents an experiment in selective privacy for low energy and scalability:

**Privacy assessment:** to ascertain where security functions can be removed safely. We informed our design by the use of a Data Protection Impact Assessment (DPIA), typical of General Data Protection Regulation (GDPR) auditing.
**Time analysis:** of attributes for securing payloads using hardware acceleration. Timing security attributes as separate entities demonstrated the consumption differences with privacy on and off.
**Power analysis:** of the same security attributes to discover the lightest use of energy. Payload security with selective privacy demonstrated an energy saving of up to 81% compared with the same TLS implementation.

The ESP32 microcontroller was used to demonstrate the energy consumption of the design, acting as the edge device, while a Raspberry Pi posed as a gateway controller to the cloud.

The rest of this paper features the background problem as discussed in part II, related works of distributed and lightweight security are discussed in part III, the design in part IV, part V is testing, and the research is concluded with considerations of further work in part VI.

## 2. PROBLEM BACKGROUND

This section discusses the problems with integrating TLS into IoT networks, specifically the smart-farm aquaponics domain. The problem is divided into three areas; false perception of threat, unnecessary privacy, and inappropriate ciphers.

### 2.1 False perception of threat

Perhaps the biggest challenge of widespread security integration is overcoming the notion that a small, lightweight, IoT-enabled network such as one powered by BLE, is subject to threat. In addition to the 'unlikeliness' that a smart-farm will be subject to attack, the centralised nature of TLS requirements is demanding for an edge device such as the ESP32. TLS configuration is complex in structure, high in energy consumption, and promotes unnecessarily stringent protection levels. Certificate Authorities (CA's), Registration Authorities (RA's), and Verification Authorities (VA's), typical of an independent network security infrastructure present vulnerabilities and implementation challenges for IoT (Höglund *et al.*, 2020). Without such an infrastructure, employing a third-party for the same centralised functions presents a different set of challenges, such as trust, downtime, and the requirement of the internet whereas edge devices could otherwise operate on lighter protocols such as BLE alone.

With fewer restrictions and structural requirements than TLS, security in general would be a more realistic consideration. Attacks on protocols such as BLE, LoRaWAN and MQTT are emerging - and security is maturing slower than our dependence on them (Anand *et al.*, 2020). Absence of security provisions has not gone unnoticed either, with attacks under development towards BLE (Pallavi and Narayanan, 2019), LoRaWAN (Ingham, Marchang and Bhowmik, 2020), and MQTT (Firdous *et al.*, 2017).

### 2.2 Unnecessary privacy

HTTP Security (HTTPS) employs TLS. This is a predefined set of rules including Confidentiality (privacy), Integrity, and Authentication (CIA), functions. Client and server devices traditionally employ TLS to negotiate a ciphersuite from a list, proceeding to exercise the full CIA protection on each message. Such enforcement is not good for constrained devices; the protection is largely overkill (Pérez Goya, 2015), and for many of the messages within smart-farm activity, content is sensor readings - predictable in nature. Traditionally, the general consensus of an adequate security design includes the CIA (Yin *et al.*, 2020), (Noor and Hassan, 2019). This triad demonstrates a construct that TLS has thrived on. Confidentiality as privacy of message content, so that only the recipient can view it. Integrity is the guaranteed non-tampering, so that it has not changed in transit. Availability is so that the content is complete, and from the alleged party sending it. Where the use of the CIA triad has provided a strong application in traditional TLS, the requirements for a full CIA composite could now be to the detriment of energy-conscious IoT solutions such as smart-farming.

### 2.3 Inappropriate ciphers

Security attributes available for the ESP32 are not limited to TLS completely, since the functions are used for CertificateLess Signature (CLS), schemes such as using Digital Signature Algorithms (DSA), Blockchain and Advanced Threshold Systems (ATS) which use SHA. Thus, the board hosts hardware acceleration and a correlating library which alludes to TLS implementation, but not contracted to it.

Therefore, the functions RSA (asymmetric encryption), AES (symmetric encryption), SHA (compression functions), and RNG (Random Number Generation), provide standards recognised by NIST and FIPS, and that might be used as the basis of a novel, IoT-oriented security model.

The challenge is to discover the appropriate, or least-consumptive cipher configurations in terms of energy to maximise the battery life of the edge devices. Whilst doing so, a good standard of security can be achieved comparable to TLS, but abiding by the same regulations.

## 3. RELATED WORK

Below we break down the challenges into increasing battery life, security attributes, and topology.

### 3.1 Increasing battery life

(Weghorn, 2013) previously discussed the use of BLE and ANT+ in personal sports devices as a much lighter alternative to Bluetooth Classic (Weghorn, 2015). Shortly after, such devices were proven to have the ability of energy harvesting for power storage within superconductors (Weghorn, 2017). Energy harvesting by other means, such as solar, have also been used in conjunction with the storage capacity of superconductors (Elahi *et al.*, 2020).

In climates with long daylight hours and intense sunlight, solar harvesting would be viable for self-perpetuating BLE-powered networks - as would vehicular applications where devices are in regular motion. However, the smart-farm is a static model, and not guaranteed to survive limited sunlight provisions such as in the UK.

Other methods for enhancing the longevity of devices include using lighter protocols and applying security to them (Aloufi and Alhazmi, 2020), reducing clock speed (Suárez-Albela *et al.*, 2018), and lightweight hash functions (Dhanda, Singh and Jindal, 2020), and ciphers such as SIMON (Alassaf *et al.*, 2019).

### 3.2 Security attributes

Perhaps exclusively to IoT smart-farming, the opportunity to differentiate between security and privacy could be of enormous contribution to the longevity of all protocols such as BLE (Zhang *et al.*, 2020).

Integrity and availability (or authenticity), are paramount for every message sent - but perhaps not so much for privacy. For IoT applications, the requirement for message protection may present a paradigm shift for the traditional CIA attributes, where there is an absence of private data (Pivoto *et al.*, 2018). Traditional security applications between HTTP-connected devices protected banking and shopping transactions, where most, if not all data featured highly sensitive exchanges. Such information has demonstrated enormous consequences in the absence of rigorous privacy, ie, fraud. However, in a typical smart-farm domain such as aquaponics or hydroponics, most of the messages from the client to the server are benign, predictable, and useless to all but the operator (Ruengittinun, Phongsamsuan and Sureeratanakorn, 2017). It is important that the value of such temperature and pH-level reading remain unaltered during transition (integrity), and that the origin of the data is verified (authentication). Beyond those two requirements, confidentiality for sensor readings is largely unnecessary.

The General Data Protection Regulation (GDPR), was an EU directive that came into force in May 2018. The GDPR proposed a series of subject rights and regulations to control the processing of personal data, introducing privacy as a separate topic to security, but with shared connotations and some overlap. Whereas security included the traditional CIA attributes, privacy was introduced as data protection under anonymisation, tokenisation, or pseudonymisation - data could be viewed by anyone, but only be of use to those who could interpret it. The difference between interpretation and encryption is a key; security models require keys to decrypt ciphertext into plaintext, and data under privacy protection does not.

Anonymisation, for anonymous data sharing, has demonstrated optimisation whilst retaining privacy (Sun *et al.*, 2011). Tokenisation has demonstrated lightweight solutions for authentication in IoT networks (Dammak *et al.*, 2019), and towards decentralisation to accommodate scalability (Naveed Aman *et al.*, 2019), but lack coverage in message security specifically. In the context of a smart-farm application domain, authenticating devices and sending messages between them are two separate studies; this paper discusses the latter. Finally, pseudonymisation has been successfully demonstrated in medical applications (Darwish, Nouretdinov and Wolthusen, 2018), incorporating query-based privacy models on distributed storage.

Amongst the recommendations and corresponding actions for GDPR compliance (Brodin, 2019), the Data Protection Impact Assessment (DPIA), is a framework for ascertaining the risks associated with data leakage. The nature of an application has a significant influence on how privacy should be modelled (Elliot *et al.*, 2018); confidentiality requirements can be addressed herewith.

### 3.3 Security topologies

Centralised models are those in which a server, or possibly a server and a third-party hold accountability for some or all the network's security authorities, such as TLS. Decentralised models such as Blockchain exist without a central authority to govern them - but rather distributed trust and reputation systems. Distributed systems in general are considered preferable to centralised models for scalability, such as the IoT (Marquez et al, 2015).

Blockchain has demonstrated privacy through anonymisation (de Haro-Olmo, Varela-Vaca and Álvarez-Bermejo, 2020), and the same privacy capacity of voting systems centuries old (Tarasov and Tewari, 2017). Blockchain has a broad range of applications, notably crypto currency (Ben Sasson *et al.*, 2014), and IoT data sharing applications using distributed ledgers, such as Iota (Silvano and Marcelino, 2020).

As an example of a Distributed Ledger Technology (DTL), Iota provides data exchange via a 'Tangle'; a reputation system by which transactions are authorised by validating two pre-existing ones. Therefore, the reputation, and enabling of data transfer, is validated as part of the network by another user. Each additional IoT device relies on the verification of pre-existing ones, providing scalability, access management, authentication and integrity in a continuously developing network (Novo, 2019), (Xu *et al.*, 2018).

Characteristically, Blockchain lends itself to the chaotic and large scale nature of IoT (Righi et al. 2018), but with a security caveat of trusting all connected devices (Kim *et al.*, 2019). Additionally, Blockchain introduces a storage and processing challenge for constrained devices - accumulative hashes, and the expectations of a reputation system to process them, known as a 'hashrate' in crypto currency mining.

In summary, maintaining battery life using energy harvesting is not practical for static environments with unreliable solar, and so other means of reducing energy consumption can be exercised by changing the configuration and topology of the security model. Firstly, the confidentiality aspect can be reduced where possible, and secondly, the central model of employing a server (or third-party), for the three authorities CA VA and RA can be removed. By removing the TLS requirement of a server and ciphersuites, protocols such as BLE, LoRaWAN and MQTT can be protected using less energy-consumptive attributes available through hardware acceleration.

### 4. DESIGN

Proposed are three design concepts to resolve the problems outlined in section II, including distributed infrastructure, selective privacy, and IoT-appropriate ciphers.

### 4.1 Distributed infrastructure

The design does not propose use of a TLS channel, but on-board TLS functions using hardware acceleration to protect every message at source. The keys used are presumed to have already been shared during authentication (not part of this study).

Messages are secured as part of the procedure opening the protocol(s), so that any protocol used to communicate the message (BLE, LoRaWAN etc), can do so without a predefined ciphersuite, and thus without HTTPS. This 'plug-and-play' payload security design enables portability, and removes the stringent requirements of the TLS regulations to function.

A distributed infrastructure is proposed, where the central server is not required for security functions, but can, as any other device is able to, decrypt all messages for data handling. The removal of a central infrastructure aims to resolve the vulnerability and responsibility of a single entity undertaking all security processing - particularly as the network scales. In addition, enabling the ESP32 microcontrollers to undertake their own security processing makes the security model portable, and inclusive of any protocol used to send messages.

In addition, with simple implementation of only a short code to utilise the model, the design should promote the appeal of security and lessen user reluctance to integrate it. This would contribute towards the threat perception issue, and safeguard smart-farms in preparation for future vulnerability.

## 4.2 Selective privacy

Without adhering to the stringent requirements of employing a TLS channel, it is also possible to remove the privacy condition, or the encryption algorithm which costs IoT networks so much energy. Although not advisable in every message, as some will contain session keys or proprietary code, the majority of messages will be sensor readings. Perhaps 99% of messages will not need encryption, saving on a lot of energy that would have otherwise been wasted.

TLS enforces the full CIA triad of security attributes within each message as part of its ciphersuite conditions. As TLS develops, the number of ciphersuites becomes fewer and more protective, and as a result, less energy-efficient. Newer ciphersuites include an all-in-one-function - the Authenticated Encryption with Associated Data (AEAD). This function incorporates a cipher for encryption, a compression, or hash, function for integrity - and a Hash-keyed Message Authentication Code (HMAC), to prove where it came from. TLS is encouraging the use of Galois Counter Mode (GCM), an AEAD function containing a very sophisticated and very IoT-averse algorithm. GCM uses around three times the energy that most other ciphers do - even with manually attached AEAD functions.

The proposal for reducing the burden of unnecessary encryption whilst still fulfilling the AEAD suggestions of TLS is to combine an available cipher with a hash function and an authentication function. If this is done manually, encryption for readings from the client to the server can simply not be included. For the majority of readings from the board to the Pi, this will be the case; every sensor reading will use a hash function and an authentication key, but no encryption. The rare exception to this rule is session reset. When the device generates a new key, every week, or month, or even year, it will require encryption. Keys will be made for each board by its own board as part of the decentralised infrastructure mentioned earlier. However, because the smart-farm contains a series of fixed, static systems, the key resets will be infrequent.

## 4.3 IoT-appropriate ciphers

Finally, the design makes use of the lightest 'safe' cipher available for the ESP32 hardware acceleration. By using a cipher without the built-in AEAD function, the full range of available ciphers can be tested for the lightest energy consumption.

The ESP32 has access to a native TLS library ported for embedded devices. The mbedtls library offers the expensive AEAD function, GCM, as well as four other standalone ciphers that can be assembled into AEAD by including a hash and HMAC. By using the lightest standalone cipher available in the mbedtls library, encryption could be left out for sensor readings from the edge devices to the gateway, but included when sending keys.

Structurally, ciphers require different implementations such as data tags, Initialisation Vectors (IV), and counter offsets. Any cipher will use a symmetric encryption key accelerated by the AES hardware accelerator, and this has always been shared at the session negotiation stage before sending a message can commence. In addition, the HMAC key responsible for ascertaining the authentic origin of the sender device, will assume a new, additional session key. In TLS, the HMAC key is encrypted as part of the message, and decrypted at the recipient end to justify authenticity of the message. However, where encryption will be mostly absent, it will act as a secret key in its own right, and should be treated with the same secrecy as a regular AES encryption key where encryption is absent.

## 5. IMPLEMENTATION TESTING

Testing was performed in three parts; privacy assessment, timing analysis and power analysis.

### 5.1 Privacy assessment

We first performed a Data Protection Impact Assessment (DPIA), which considered each possible message between the gateway and the ESP32 edge device. The structure of a DPIA varies between templates, but the purpose is to calculate the risk severity of each type of data within a system, and use that risk score to address security measures (Ando, et al 2018). Complex systems holding a lot of user data in various forms, such as healthcare, require gradients of privacy measures including pseudonymisation and anonymisation. The smart-farm is simple in comparison, because there is a clear difference between the content of the systems, and that of living subjects. Security attributes for this application could be determined in a binary way; the messages were either private, and required encryption, or they were public, and they did not. This type of risk assessment is typically used by GDPR Practitioners as part of an ISO27001 audit (*ISO/IEC 27001:2013*, 2019).

Table 1. Privacy requirements of messages

| Sender | Recipient | Message content | Sensitivity | Vulnerability | Impact | Score | Risk |
|---|---|---|---|---|---|---|---|
| Pi | ESP32 | Farm control | 2 Professional | 1 Exceptional | 2 Reputation | 1.7 | High |
| ESP32 | Pi | Session keys | 3 Private | 1 Exceptional | 3 Closure | 2.3 | Critical |
| ESP32 | Pi | Temperature | 0 Public | 0 None | 0 None | 0.0 | None |
| ESP32 | Pi | Humidity | 0 Public | 0 None | 0 None | 0.0 | None |
| ESP32 | Pi | Nitrate | 0 Public | 0 None | 0 None | 0.0 | None |

Risks of a high or critical score required encryption, and risks returning low or zero scores did not. Fortunately, most of the readings from the ESP32 to the Pi did not require encryption.

### 5.2 Timing analysis

Real-time data delivery for agricultural systems and reducing latency enables good practice (Lopes and Vaz, 2019). With the lowest energy consumption possible, a good security application should also reduce latency. Timing of each attribute was assessed by including the Arduino micros() command as a component of the ESP-IDF (*IoT Development Framework I Espressif Systems*, no date). Micros() was used to measure the processing time only of the attribute, excluding variables, parameters, and all printing to the serial monitor. Each attribute was repeated in iterations of 10, 20, 30 and 40 thousand to calculate an average time for a single iteration, across various results.
The board was prepared for fair testing by ensuring that the WatchDog Timer (WDT) was disabled, as this can cause interruptions and distort readings, and that all hardware accelerations were enabled under mbedtls (ARM Limited, no date b)options. These configurations were undertaken using the menuconfig option, part of the ESP-IDF.

 A single character was defined as a byte, and text input volumes of two lengths were processed for each test. An input of 64 characters was used to demonstrate the energy required for securing a reading or other small string, and that of 512 bytes reflected a very strong key (4096 bits). A strong key length was chosen because of the intention to reset keys very infrequently, perhaps once a year.

 Firstly, integrity and authentication functions where an HMAC-SHA of sufficient strength was tested for the shortest processing time. These were SHA lengths of 224 and above. It was important to assess HMAC-SHA independently of encryption modes to demonstrate how much more energy efficient messages could be without using a cipher.

Table 2. Integrity function time analysis measured in microseconds (μS)

| Hash-keyed MAC function | 64 bytes input | 512 bytes input |
|---|---|---|
| HMAC-SHA224 | 150 | 403 |
| HMAC-SHA256 | 91 | 163 |
| HMAC-SHA384 | 102 | 165 |
| HMAC-SHA512 | 105 | 167 |

Secondly, a time assessment of the cipher modes available in mbedtls on the ESP32 (ARM Limited, no date a); GCM, CBC, CTR, CBF8 and CFB128 was undertaken. Each cipher was coupled with the same HMAC-SHA function showing the shortest processing time in results set one.

Table 3. AEAD function time analysis measured in microseconds (μS)

| AEAD function | 64 bytes input | 512 bytes input |
|---|---|---|
| AES128-GCM | 178 | 856 |
| AES128-CBC-HMAC-SHA256 | 124 | 276 |
| AES128-CTR-HMAC-SHA256 | 123 | 352 |
| AES128-CFB8-HMAC-SHA256 | 303 | 1782 |
| AES128-CFB128-HMAC-SHA256 | 121 | 350 |

In full AEAD function using the lightest integrity and authentication, CFB8 proved to be the heaviest cipher, followed by GCM. The overall lightest mode was CBC, since CTR proved to increase significantly for larger messages. Although the two ciphers could be used simultaneously in practice, the structural requirements for CBC and CTR vary, and so the use of CBC alone would simplify implementation whilst remaining the lowest consistent encryption mode.

In summary, where privacy is not required, the lightest function was HMAC-SHA256, and where privacy is required, CBC mode was a low time consumer and most consistently low.

## 5.3 Power analysis

Measurements were made at the necessary micro to milli levels for such protocols as BLE (Kamath and Lindh, 2010) using an oscilloscope for sampling (Zwerg *et al.*, 2011). Power in milliwatts (mW), was calculated by multiplying current in milliamps (mA), by voltage in millivolts (mV), with a known resistance in Ohms (Ω). The oscilloscope took the readings of the board using the digitalWrite() Arduino command to communicate the sketch over General Pin Input Output (GPIO), 21, and connecting from the ground pin to complete the circuit. As with timing analysis, integrity and availability power consumptions were assessed first, followed by the AES cipher in various modes, fulfilling the AEAD requirements with the least power-intense HMAC-SHA.

Table 4. Integrity function power analysis measured in milliwatts (mW)

| Hash-keyed MAC function | 64 bytes input | 512 bytes input |
|---|---|---|
| HMAC-SHA224 | 156 | 157 |
| HMAC-SHA256 | 155 | 155 |
| HMAC-SHA384 | 156 | 157 |
| HMAC-SHA512 | 154 | 154 |

Table 5. AEAD function power analysis measured in milliwatts (mW)

| AEAD function | 64 bytes input | 512 bytes input |
|---|---|---|
| AES128-GCM | 158 | 158 |
| AES128-CBC-HMAC-SHA256 | 155 | 155 |
| AES128-CTR-HMAC-SHA256 | 155 | 155 |
| AES128-CFB8-HMAC-SHA256 | 155 | 155 |
| AES128-CFB128-HMAC-SHA256 | 156 | 156 |

Finally, we summarise the energy consumptions of each attribute by multiplying time (μS) by power (mW) to conclude energy in seconds, Joules (J).

Table 6. Integrity function energy consumption measured in Joules (J)

| Hash-keyed MAC function | 64 bytes input | 512 bytes input |
|---|---|---|
| HMAC-SHA224 | 23.4 | 63.271 |
| HMAC-SHA256 | 14.105 | 25.265 |
| HMAC-SHA384 | 15.912 | 25.905 |
| HMAC-SHA512 | 16.17 | 25.718 |

Table 7. AEAD function energy consumption measured in Joules (J)

| AEAD function | 64 bytes input | 512 bytes input |
|---|---|---|
| AES128-GCM | 28.124 | 135.248 |
| AES128-CBC-HMAC-SHA256 | 19.22 | 42.78 |
| AES128-CTR-HMAC-SHA256 | 19.065 | 54.56 |
| AES128-CFB8-HMAC-SHA256 | 46.965 | 121.21 |
| AES128-CFB128-HMAC-SHA256 | 18.876 | 54.6 |

CBC in full AEAD mode demonstrated the best potential for battery longevity.

Where readings do not require privacy, the energy consumption is as low as 14 Joules, and if exchanging an instruction set for farm control without sensitive data, 25 Joules per second. Compared with the TLS recommendations of GCM at 28 and 135 Joules per second respectively, the energy saving without privacy can be between 50% for a small reading, and over 81% for a large exchange of 512 bytes.

If that data did require encryption and the full AEAD capacity such as a key, the difference between GCM and CBC in AEAD mode in this comparison is over 32% for smaller exchanges such as readings, and 68% lighter for key exchanges or sensitive instruction sets.

## 6. CONCLUSION

The contributions and further work of this research are concluded below.

**Distributed infrastructure.** Our security model uses microcontroller hardware as a source of security from dedicated board functions, and without supervision of a central server. This decentralised, distributed payload security model allows multiple protocols without protocol protection.

**Selective privacy.** By not requiring a TLS channel, the design proposed that messages are considered on individual merit with regards to confidentiality. Where the majority are composed of readings, benign and useless to all but the gateway, privacy can be forsaken for an energy saving of around 81%. Of course, integrity and authentication functions are provided for all messages.

**IoT-appropriate ciphers**. The design proposed full AEAD functionality for all ciphers and not just the predefined GCM recommended as part of the de facto TLS standard. Encouraged for content confirmed as high or critical risk, the alternative use of the CBC cipher still proved to be 68% more energy efficient in larger exchanges where full AEAD would be suitable - keys and instruction sets.

These savings are considerable, and could help overcome the perceived challenge of security applied to IoT-oriented smart-farm applications such as aquaponics. In addition, further work could include healthcare or smart city monitoring where anonymous data can be gathered from machines and living subjects separately.